\documentclass[prl,twocolumn,showpacs]{revtex4}
\usepackage{graphicx}
\usepackage{amssymb}
\usepackage{amsmath}
\usepackage{amsbsy}
\usepackage{color}
\usepackage{dcolumn}
\usepackage{hyperref}

\begin{document}

\title{Correlations between the final momenta of electrons and their initial phase-space distribution in the process of tunnel ionization}

\date{\today}

\author{V. Ayadi} 
\affiliation{MTA "Lend\"ulet" Ultrafast Nanooptics Group, Wigner Research Centre for Physics, Konkoly-Thege 
M. \'ut 29-33, H-1121 Budapest, Hungary}

\author{P. F\"oldi} 
\affiliation{Department of Theoretical Physics, University of Szeged, Tisza Lajos k\"or \'ut 84, H-6720 Szeged, Hungary}

\affiliation{ELI-ALPS, ELI-HU Non-profit Ltd., Dugonics tér 13, H-6720 Szeged, Hungary}

\author{P. Dombi} 
\affiliation{MTA "Lend\"ulet" Ultrafast Nanooptics Group, Wigner Research Centre for Physics, Konkoly-Thege 
M. \'ut 29-33, H-1121 Budapest, Hungary}

\affiliation{ELI-ALPS, ELI-HU Non-profit Ltd., Dugonics tér 13, H-6720 Szeged, Hungary}

\author{K. T\H{o}k\'{e}si} 
\affiliation{Institute for Nuclear Research, Hungarian Academy of Sciences (ATOMKI), H-4026 Debrecen, P.O. Box 51, Hungary}

\begin{abstract}
We present both full quantum mechanical and semiclassical calculations of above threshold ionization (ATI)
of a hydrogen atom in the tunneling regime by a few-cycle linearly polarized infrared laser pulse. As a quantum treatment, 
we applied the direct integration of the time dependent Schrödinger equation (TDSE). In the semiclassical approximation (SCA), it is assumed that 
wavepacket propagation in the post-tunneling process can be well described within the classical framework. 
With these two methods, we analyze the similarities and deviations for ionization of the hydrogen atom.
We found that the 3 dimensional semiclassical method can describe reasonably well the momentum 
correlation pattern of the ATI peaks. We also show good agreement between 
the results obtained by TDSE method and the semi-classical method. Furthermore, with the semiclassical approximation
we clearly identify and separate the regions in momentum distributions of the ejected electrons according to initial conditions. 
We illustrate the corresponding regions with typical electron trajectories.

\end{abstract}


\maketitle

\section{Introduction}

Understanding the ionization process during atomic collisions is fundamental both from the 
experimental and theoretical points of view. Especially, it is a challenging theoretical task  to 
describe the ionization cross sections near the threshold region. It was shown that the interaction of a short, 
few-cycle infrared laser pulse with an atom characterized initially by superposition of two stationary states 
exhibits strong signatures of atomic coherence \cite{Ay}. Along this line, we calculate the above threshold ionization 
(ATI) spectra and the angular distribution of electrons ejected from the hydrogen atom in the tunneling regime for the ground state. 

The calculations to be presented below use both full quantum mechanical and semiclassical methods. 
We applied the direct integration of the time dependent Schrödinger equation (TDSE) and the
semiclassical approximation (SCA) \cite{Li, Hu, Fey}. The latter approach is similar to the Classical Trajectory Monte Carlo (CTMC)
method, and it is based on the inclusion of the classical phase information of the motion. Over the past years SCA has widely been used
for investigation of laser-atom collision, partly because it is much simpler than any other quantum treatment of the problem and it holds also
the possibility of the visualization of the electron trajectories in a certain momentum map.
This fact is true even when a large number of electron trajectories have to be determined,
typically 100 million, for the accurate description of the tunnel ionization.

Although the full quantum mechanical treatment is the "ultimate" way of describing atomic processes, sometimes it does not provide an intuitive picture. 
In other words, all physically relevant (or meaningful) questions could in principle be answered in terms of quantum mechanics,  but the solution of the TDSE (especially when it can only be done by numerical means) may not reveal the physical mechanisms responsible for the observable effects. This is the point, when an appropriate semiclassical method, that is proven to be able to deliver measurable results close to the predictions of quantum mechanics,  can be very useful. In this case, using the semiclassical model, we can ask and answer questions that -- in extreme cases -- can even be incompatible with the laws of quantum mechanics, but still add important contribution to our understanding of the investigated physical process. In this way, the semiclassical model helps developing a clear, physical picture -- which, finally, should clearly be compatible also with quantum theory.  

Along this line,  in the first part of this paper, we show that the full quantum mechanical model and the semiclassical approach provides very similar results for the momentum distribution of the liberated electrons in the process of tunnel ionization. Based on this, later on we focus on the semiclassical model and investigate the correlations between the initial phase-space positions of the (bound) electrons and their final momentum distribution. This allows us to assign an intuitive picture to the ionization mechanism, by identifying the initial conditions that correspond to the well separated interference maxima in the final momentum distributions.
Atomic units are used throughout the paper unless indicated otherwise.

\section{Theory}

In our simulations we use a few-cycle linearly polarized infrared laser pulse.
The polarization vector of the field is fixed along the $z$ axis as excitation source. The vector potential of the external
laser pulse is assumed to be polarized in the $z$ direction, so the only non-vanishing 
component is
\begin{equation}
A_z(t)=-(F_0/\omega)\sin^{2}\left(  {{\pi t}/{\tau}}\right) \sin(\omega t + \varphi
_{\mathrm{CEP}}) \, , 
\label{vectpot}
\end{equation}
in the Coulomb gauge, where the $\sin^{2}$ envelope function  is assumed to be zero when $t<0$ or
$t>\tau$ \cite{BandraukPhysRevA04}.

The time dependence of electric field of the few-cycle lase pulse 
can be written as
\begin{subequations}
\begin{align}
F(t)&=F_1(t)+F_2(t) \, , \label{EfieldA}\\
&\quad \textrm{with} \nonumber \\
F_1(t)&=F_{0}\sin^{2}\left(  {{\pi t}/{\tau}}\right)  \cos(\omega t+\varphi
_{\mathrm{CEP}}) \, , \label{EfieldB} \\
F_2(t)&=\frac{\pi F_{0}}{\omega\tau}\sin\left(2{{\pi t}/{\tau}}\right)  \sin(\omega t+\varphi
_{\mathrm{CEP}}) \, , \label{EfieldC}
\end{align}
\end{subequations}
where $\omega$ is the angular frequency corresponding to a central wavelength of $800 \ \mathrm{nm}$ and 
$\tau$ is the temporal length of the $n$ cycle pulse, which is given by $\tau = 2\pi n / \omega$. 
Additionally, unless stated otherwise, we use $\tau=21.4\, \mathrm{fs}$ [corresponding
to $7.8\,\mathrm{fs}$ intensity full width at half maximum (FWHM) (8 cycle)]
and $F_{0}=25\, \mathrm{GV/m}$ is the peak field strength.
These parameters are experimentally achievable using current femtosecond amplifiers
\cite{Mourou}. We also assume $\varphi_{\mathrm{CEP}} = 0$. 

\begin{figure}[ht!]
\begin{center}
\includegraphics[width=9cm]{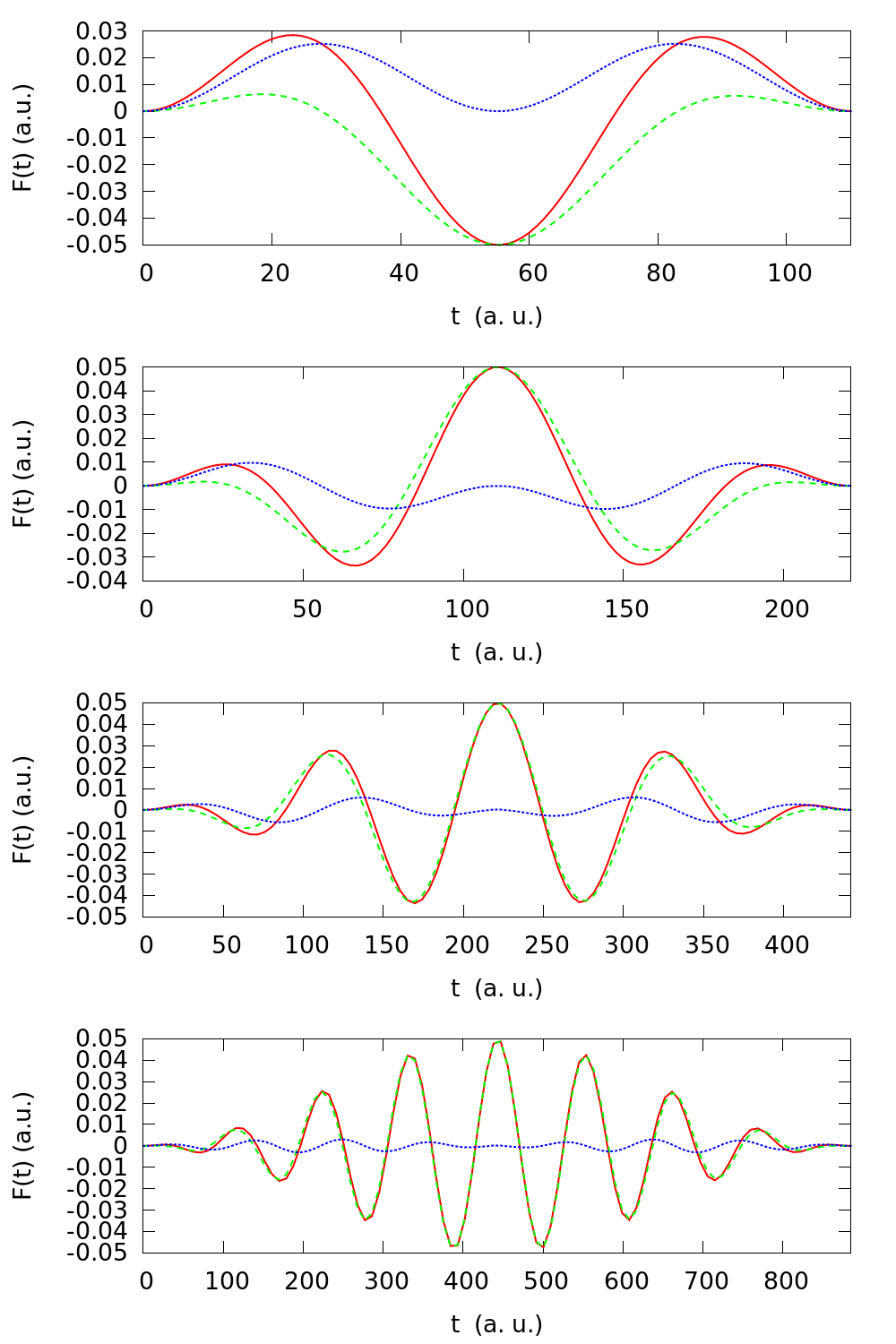}
\end{center}
\caption{The shape of the typical pulses in time domain. The parameters of the electric field are described by Eq. (\ref{EfieldA}-\ref{EfieldC}) are as follows: 	
$F_0 = 0.05$ a.u. and $\omega = 0.0567$ a.u. a) $n=1$ cycle pulse ($\tau=110$), b) $n=2$ cycle pulse ($\tau=221$), c) $n=4$ cycle pulse ($\tau=442$),
d) $n=8$ cycle pulse ($\tau=885$). The solid line denotes the total field strength $F(t)$, the dashed line denotes $F_1(t)$ term and the dotted lines corresponds to the $F_2(t)$ correction term.
}
\label{fig1}
\end{figure}

As an example, figure \ref{fig1} shows the shape of a typical pulse used in the present calculations.

\subsection{Time dependent Schr\"odinger equation}

We use atomic units and solve the time dependent Schr\"{o}dinger equation
(TDSE) numerically in the coordinate representation:
\begin{equation}
i\frac{\partial}{\partial t}\psi(\mathbf{r},t)=H(t)\ \psi(\mathbf{r},t),
\label{TDSE}%
\end{equation}
where the Hamiltonian takes the form
\begin{equation}
H(\mathbf{r},t)=H_{0}(\mathbf{r})+H_{\mathrm{I}}(\mathbf{r},t)=\left[  -\frac{\Delta
}{2}-\frac{1}{r}\right] + F(t)z. \label{Ham}%
\end{equation}
{Note that the position vector $\mathbf{r}=(x,y,z)$ corresponds to the
relative (electron-nucleus) coordinates, i.e., we are working in the center of
mass frame. The light-atom interaction described by $H_{\mathrm{I}}$ }${=}F(t)z$ {is
written using the dipole approximation (which can be shown to be valid in the
parameter range considered here \cite{BandraukJPhysB13, ReissJPhysB14}).
}

Note that various numerical methods can be used for the solution of the
TDSE as a partial differential equation, for example: the method of lines,
split step Fourier, etc. For our purposes, the most efficient approach
was found to be based on spherical harmonics expansion
\cite{Krause92}. That is, we use spherical coordinates and write $\psi$ as
\begin{equation}
\psi(r,\theta,t)=\sum_{\ell=0}^{\ell_{\mathrm{max}}}\frac{\Phi_{\ell}(r,t)}%
{r}Y_{\ell}^{0}(\theta), \label{psiBase}%
\end{equation}
where, due to the cylindrical symmetry, no summation with respect to $m$ appears:
we can restrict our calculations to $m=0.$ (In other words, there is no
$\varphi$ dependence.) Note that Eq.~(\ref{psiBase}) means a separation of the
$1/r$ dependence of the wave function, resulting in the following equations
for $\Phi_{\ell}(r,t)$  and $\hat{\psi}=r\psi:$
\begin{equation}
\hat{H}_{0}{\Phi_{\ell}(r,t)}=\left[  -\frac{1}{2}\left(  \frac{\partial^{2}%
}{\partial r^{2}}-\frac{\ell(\ell+1)}{r^{2}}\right)  -\frac{1}{r}\right]  {\Phi
_{\ell}(r,t)}, \label{H_0}%
\end{equation}%
\begin{equation}
i\frac{\partial}{\partial t}\hat{\psi}(\mathbf{r},t)=\left[  \hat{H}_{0}%
+\hat{H}_{\mathrm{I}}(t)\right]  \ \hat{\psi}(\mathbf{r},t). \label{TDSE2}%
\end{equation}
Since $H_{\mathrm{I}}$ does not contain derivatives with
respect to $\mathbf{r},$ the interaction $\hat{H}_{\mathrm{I}}=H_{\mathrm{I}}$ contains only the
$z$ coordinate, the nonzero matrix elements of which are well known, and read
\begin{equation}
\langle Y_{\ell}^{0}|\cos\theta|Y_{\ell+1}^{0}\rangle={\frac{\ell+1}{\sqrt
{(2\ell+1)(2\ell+3)}}}. \label{c_l}%
\end{equation}
The radial equation above is discretized by a special finite difference (FD)
scheme, which was presented in details in Ref.~\cite{MullerEfficient} and
relies on the alternating direction implicit (ADI) method \cite{PeacemanADI}.

For our calculations $\ell_{\mathrm{max}}$ was chosen to be $100$ and the radial
grid consisted of $10000$ points. The size of the computational grid was $1000$ atomic units
($\approx 53$ nm) in both directions. These figures were found
to be sufficient, since  the populations of the states close to the
maximal $\ell$ and $r$ values were always negligible in our calculations, i.e.,
there were no numerical artifacts
due to "reflections" at the edges of the grid (we used mask function).

The 2 dimensional momentum distributions for the TDSE were calculated
from the wave function (\ref{psiBase}) similar as in
\cite{Bandrauk2005}. First we are projecting the
asymptotic part out $\psi_\mathrm{out} (r, \theta)$
on the plane wave $\psi_f(p, \theta) = (1/(2\pi)^{3/2}) \exp (\mathrm{i}\mathbf{p}\cdot\mathbf{r})$ 
first, then we calculated the probability amplitude a $a(\mathbf{p})$ using
\begin{equation}
 a(\mathbf{p}) = (1/(2\pi)^{3/2}) \int \exp (\mathrm{i}\mathbf{p}\cdot\mathbf{r}) \psi_{\mathrm{out}}(r, \theta)r^2 dr d\Omega \, .
\end{equation}
Finally we can write the probability amplitude of final momentum eigenstate by $\vert a(p, \theta) \vert^2$
\begin{equation}
a(p, \theta) = \frac{1}{(2\pi)^{3/2}} \sum_{\ell=0}^{\ell_{\mathrm{max}}} \sqrt{\frac{2}{\pi}} (-\mathrm{i})^{\ell} Y_0^{\ell}(\theta) \int j_{\ell}^{*}(pr)\Phi_{\ell, \mathrm{out}}(pr)r\, dr
\end{equation}
where $j_{\ell}$ denotes the spherical Bessel functions \cite{Stegun}. The equations of motions
were integrated by explicit embedded Runge-Kutta Prince-Dormand method \cite{RKPD}.

\subsection{Semiclassical approach}

For any two-step step semiclassical model we first need initial conditions (starting point and initial velocity)
for electron trajectories. To obtain a starting point for a trajectory we first need to determine
the tunnel exit point, which can bee found by studying the Schr\"odinger equation in the static limit
\begin{equation}
 \left(-\frac{1}{2}\nabla - \frac{1}{r} + F z\right)\psi = - I_p \psi ,
\end{equation}
where $I_p$ is the ionization potential. It is convenient to introduce parabolic coordinates, since
the equation takes a separable form \cite{Landau3}. For obtaining the exit point, the following
equation is relevant:
\begin{equation}
 U_2(\eta) = -\frac{\beta_2}{2\eta} - \frac{m^2-1}{8\eta^2} - \frac{1}{8}F\eta, 
\end{equation}
and the exit point can calculated by solving 
\begin{equation}
 U_2(\eta) = -\frac{I_p}{2}, \label{condEta}
\end{equation}
where $\beta_2$ is a separation constant and $z$ can be aproximated as $z = -\frac{1}{2}\eta$. For the initial state 1s the parameters are given by $I_p = 1/2$, and $\beta_2=1/2$.

This equation can be solved analytically, by Cardano's method [citation], to this end 
we rearrange the equation ($\ref{condEta}$), than we obtain:
\begin{equation}
\eta^3-\frac{2I_p}{F}\eta^2+\frac{4\beta_2}{F}\eta+\frac{1}{F}= 0
\end{equation}
the largest root of the equation is given by:
\begin{equation}
\eta_0 = \frac{2I_p}{3F}+2\sqrt{-\frac{p}{3}}\cos\left\lbrace\sqrt{-\frac{3}{p}}\arccos\left(\frac{3q}{2p}\right)\right\rbrace
\end{equation}
\begin{equation}
p=\frac{4\beta_2}{F}-\frac{1}{3}\left(\frac{2I_p}{F}\right)^2 \qquad q=\frac{1}{F}+\frac{8I_p\beta_2}{3F^2} - \frac{2}{27}\left(\frac{4\beta_2}{F}\right)^3.
\end{equation}

For describing the first step of our calculation we need to describe the tunneling mechanism. This is given by
\begin{equation}
 w(F, v_{\perp}) = w_0(F(t))\frac{v_{\perp}}{F\pi}\exp{\left(-\sqrt{2I_p}\frac{v_{\perp}^2}{F} \right)} \, ,
\end{equation}
the velocity distribution of the electrons in the tunneling regime, where
\begin{equation}
 w_0(F) = \frac{4}{F}\exp{\left( - \frac{2\sqrt{(2I_p)^{3}}}{3 F(t)}\right)}
\end{equation}
and we also assume that the $v_{\parallel}$ component of the initial velocity is $0$ \cite{Li}.
The above formula can be derived by the Landau-Dykhne adiabatic approximation \cite{Delone91, Delone98}.

For the second step of our model, we have to evolve the electrons ''born'' in the first step
according to the Newtonian equations of motion 
\begin{equation}
 \ddot{\mathbf{r}} = -\frac{\mathbf{r}}{r^3} - \mathbf{F}(t)
\end{equation}
and we also have to assign a phase by the formula:
\begin{equation}
 \Phi(t_0,v_{\perp})=-\int_{t_0}^{\infty}\left[\frac{v^2(t)}{2}-\frac{2}{r(t)} + I_p\right] dt \, ,
\end{equation}
which can derived by investigating the lowest order contribution of the Feynamn path integral \cite{Fey},
a similar calculation can be found in the appendix of \cite{Li}. Finally we have to calculate the asymptotic 
velocities of the electron according to Kepler rules: 
\begin{equation}
\mathbf{p} = p\frac{p(\mathbf{L}\times\mathbf{A})-\mathbf{A}}{1+p^2L^2}
\end{equation}
\begin{equation}
\frac{p^2}{2}= \frac{p_f^2}{2}-\frac{1}{2}\qquad \mathbf{L}=\mathbf{r}_f\times\mathbf{p}_f \qquad \mathbf{A}=\mathbf{p}_f\times\mathbf{L}-\frac{\mathbf{r}_f}{r_f} \, ,
\end{equation}
and then coherently sum the energy bins of the 2 dimensional distribution.

\section{Results and discussion}

\begin{figure}[ht!]
\begin{center}
\includegraphics[width=9cm]{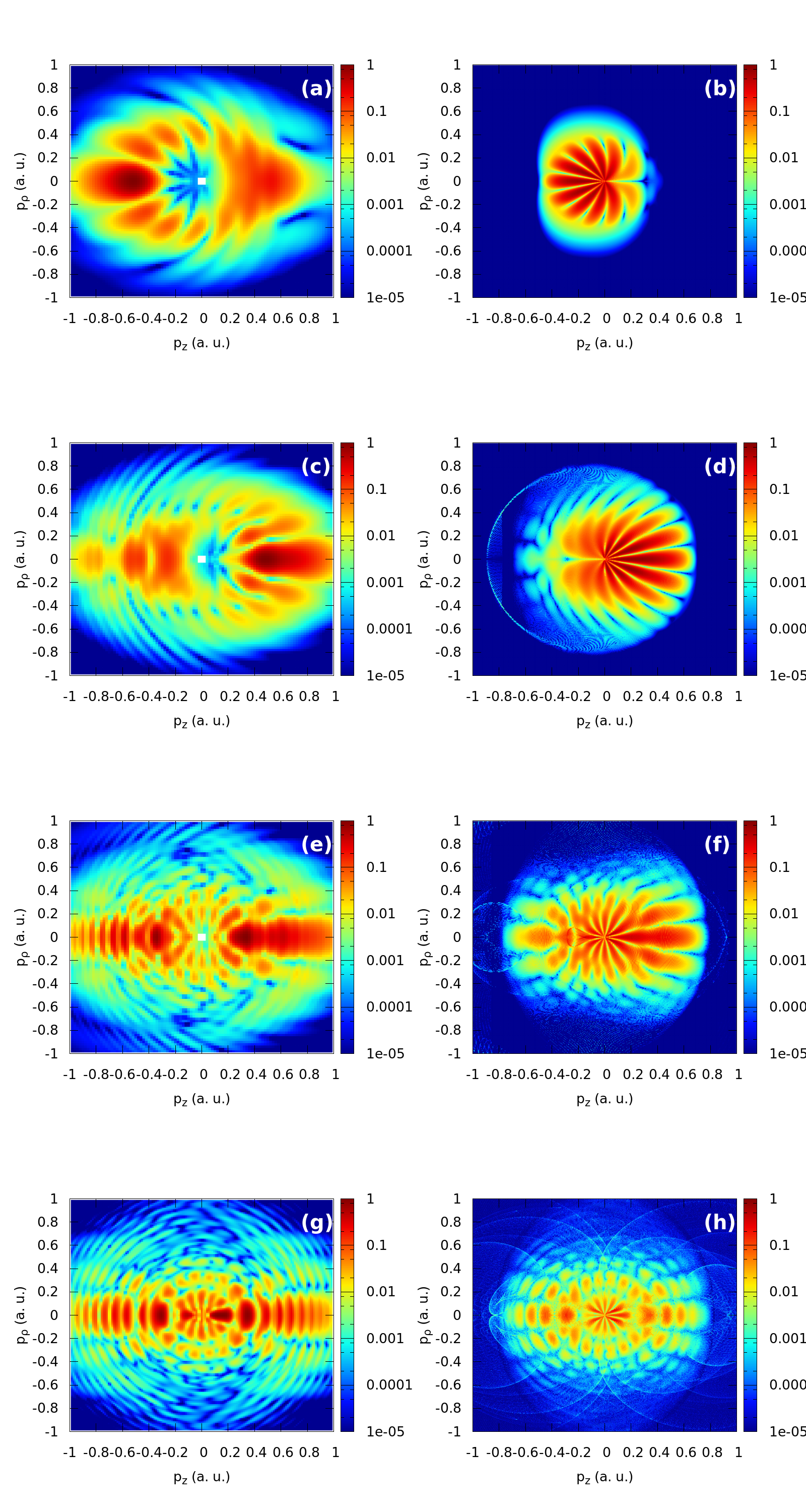}
\end{center}
\caption{Ionization probability densities for the H atom as a function of the electron parallel and perpendicular momentum 
measured from the polarization vector, which coincides with the $z$ axis. The electric field is defined by Eq. \ref{EfieldB}
First column TDSE, second column SCA. First row: 1 cycle pulse ($n=1$), second row: 2 cycle pulse ($n=2$), third row: 4 cycle pulse ($n=4$), 
fourth row: 8 cycle pulse ($n=8$).
}
\label{fig2}
\end{figure}

We determine the ionization probability densities of the H atom as a function of the electron parallel ($p_{\parallel}$) 
and perpendicular ($p_{\perp}$) momentum 
measured from the polarization vector. Figure \ref{fig2} shows the above threshold ionization probability densities for the H atom as a function of the number of cycles of the excited infrared laser pulse.
We found very strong forward-backward asymmetry in the momentum distribution at low $n$ values. 
At the same time, the interferecnce structure is much more rich for TDSE than for SCA, but still we can identify the
corresponding fan-like structure in SCA distributions also. We note, that we found significantly more electrons in the backward 
direction for $n=1$ than at forward direction. In contrast for $n=2$ most electrons are emitted in the forward direction. 

\begin{figure}[ht!]
\begin{center}
\includegraphics[width=9cm]{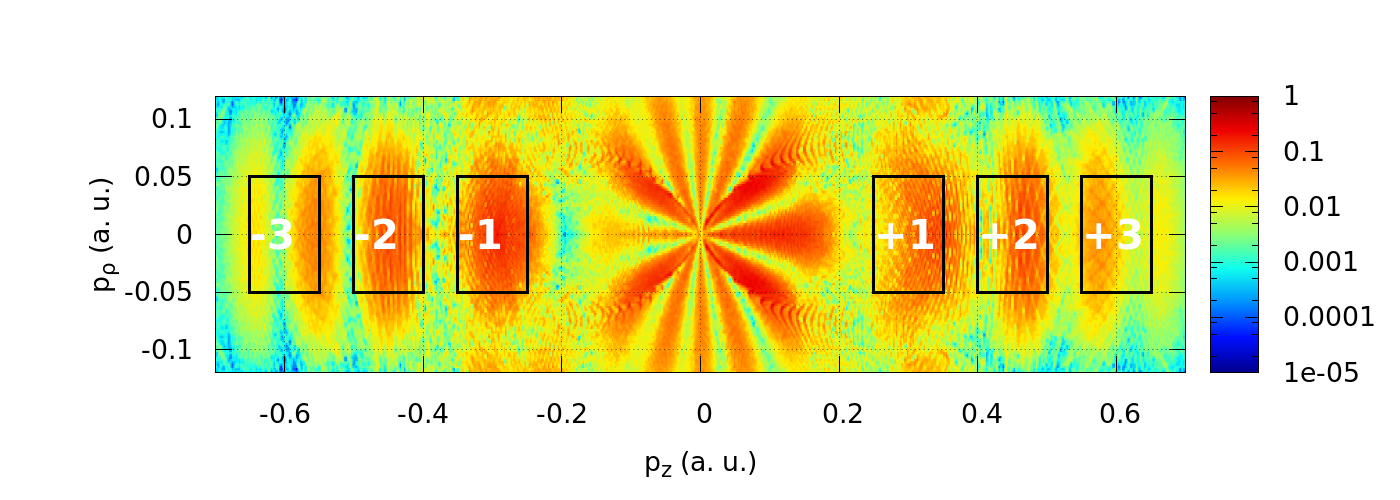}
\end{center}
\caption{The magnification of Fig. 1h for $p_{\perp}$ $<$ 0.1 a.u, which represents
the electron momentum distribution after the $n=8$ cycle pulse. The boxes on denote 
the six regions that will be studied.
}
\label{fig3}
\end{figure}

At higher $n$ the strong forward-backward asymmetry becomes less pronounced or completely disappear.
This observation is valid for both TDSE and SCA. 
The overall agreement between TDSE and SCA is improved also with increasing the pulse cycle number.
In the following investigation we will focus on the case of pulse number $n=8$. 
Semiclassical simulations have many advantages. First, these methods can be easily applied to systems 
with nontrivial geometries. Second, semiclassical simulations can help to identify the specific mechanism responsible 
for the relevant phenomena, and provide an illustrative picture of this mechanisms in terms of classical trajectories. 
According to the qualitative agreement between TDSE and SCA in Fig 1g and Fig 1h we analyse in details 
the various regions in momentum distributions of the ejected electrons according to initial conditions.
Figure \ref{fig3} shows the magnification of Fig. 1h for $p_{\perp} < 0.1$ a.u. The boxes denote the regions of 
our further investigations. The numbers in the boxes will be our reference numbers.

\begin{figure}[ht!]
\begin{center}
\includegraphics[width=9cm]{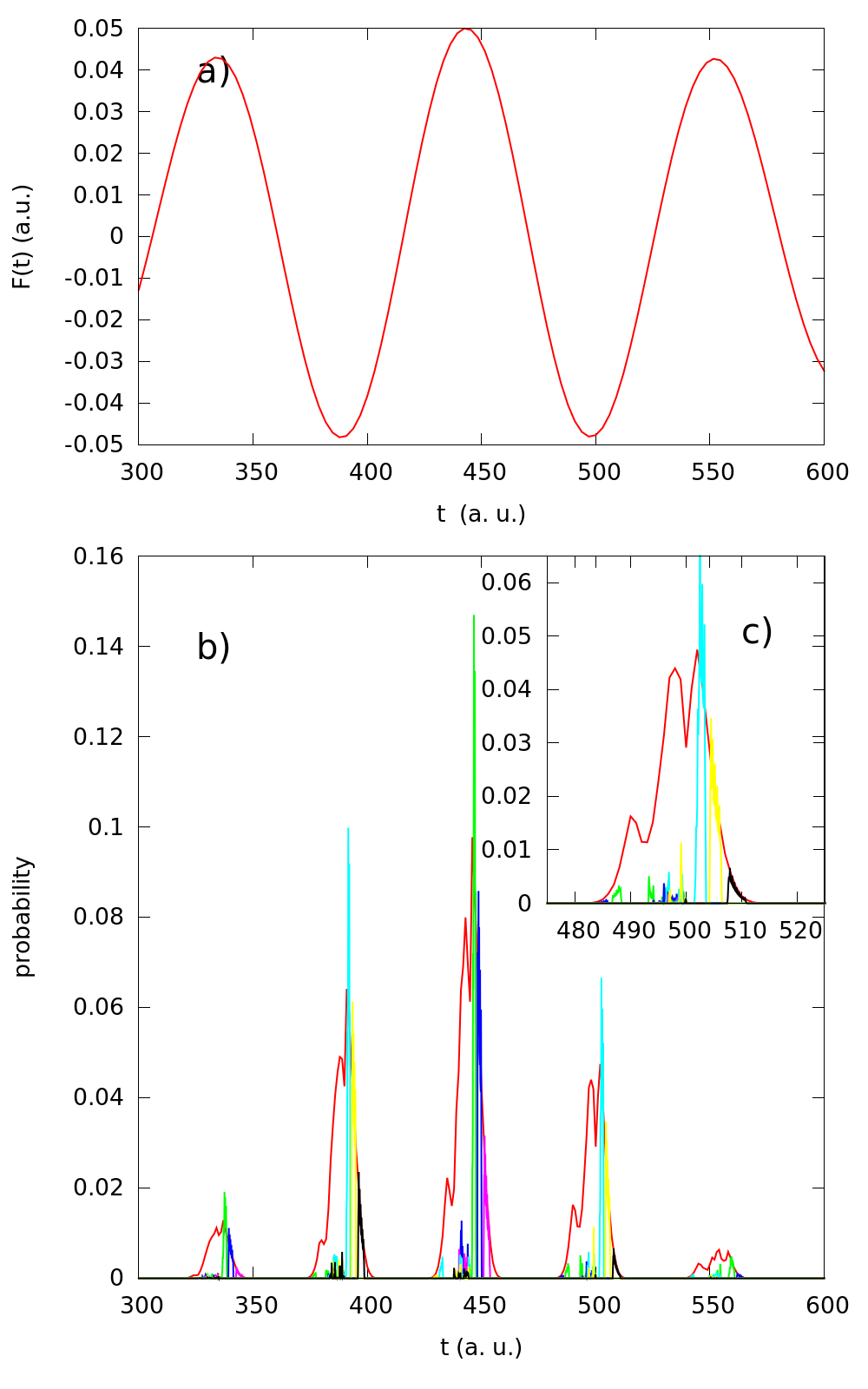}
\end{center}
\caption{The time dependence of the external  a) electric field, and b-c) the temporal distributions
of the  ''birth'' time of the electrons arriving to characteristic regions marked on $\ref{fig3}$.
The red line is proportional to the total emission, the green line corresponds to the $+1$ region, the
blue to the $+2$ region, the purple to the $+3$ region, the cyan to $-1$, the yellow to $-2$ and the black to $-3$. 
}
\label{fig4}
\end{figure}

\begin{figure}[ht!]
\begin{center}
\includegraphics[width=9cm]{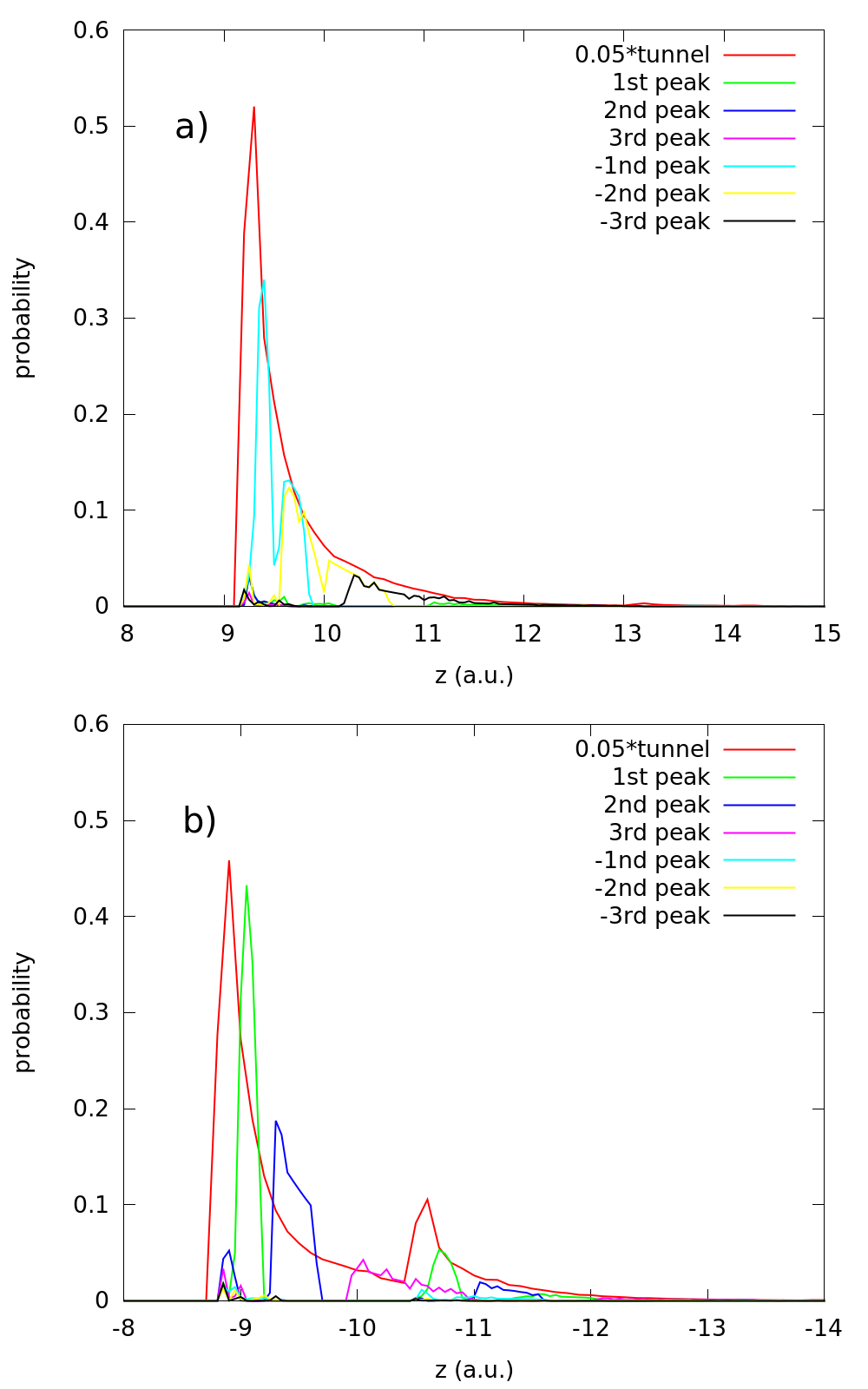}
\end{center}
\caption{The tunneling distance distribution for the electrons ''born'' on the a) right side, b) left side of the $z$ axis.
The red line is proportional to the total emission, the green line corresponds to the $+1$ region, the
blue to the $+2$ region, the purple to the $+3$ region, the cyan to $-1$, the yellow to $-2$ and the black to $-3$. 
}
\label{fig5}
\end{figure}

We sorted the events according to the certain part of $(p_{\parallel}-p_{\perp})$ (-3 $\to$ 3 in Fig \ref{fig3}) as a 
function of tunnel ionization time, initial velocities and tunnel exit points. 
Figures \ref{fig4}-\ref{fig5} show the birth time and tunnel exit point
distributions of tunneled electrons in these 6 interesting regions,
respectively.

We clearly identify and separate the regions in momentum distributions of the ejected electrons according to initial conditions. 
The separation is especially noticable for the distribution of the birth time of electrons arriving to different bumps of the momentum distribution (see the insert in Fig. \ref {fig4}).

\begin{figure}[ht!]
\begin{center}
\includegraphics[width=8cm]{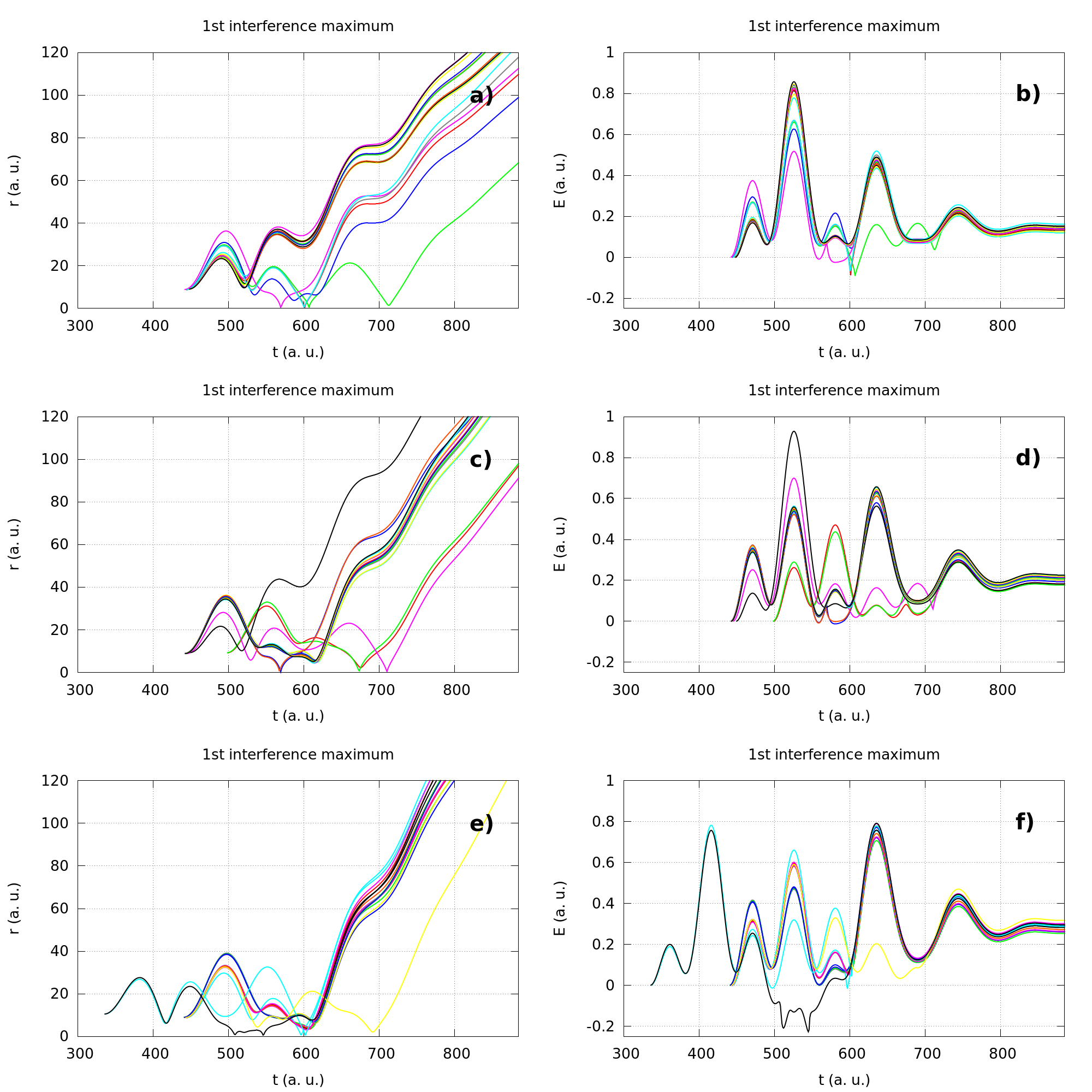}
\end{center}
\caption{Time evolution of the characteristic trajectories with the corresponding energies 
for the right side "bumps" of the $\rho$ axis a-b) first bump, c-d) second bump: e-f), third bump.
}
\label{fig6}
\end{figure}

Let us take one of the most important advantage of the classical treatment and check the typical trajectories in various regions. 
Figure \ref{fig6} shows typical classical electron trajectories in the certain regions with the corresponding energies as a function of time.
In all cases the electron trajectories are modulated by the laser field and the electron trajectories show 
oscillation, i.e., they travel back and forth from the nucleus. This is a direct consequence of the 
strong interaction with the remaining target nucleus.  
We found two completely different branch of trajectories for each region.
In one case the electrons after tunnel ionization never come back to the bound state of the target. However,
a small number of electrons return so close to the target nucleus that in short time become again bound to the target. 
This behaviour can be seen and verified by the time evolution of the electron energies.  Moreover, we can 
also see that after each close collision with the nucleus the electron can gain energy and after a 
few collisions the final electron energy is always much higher than the its energy at the birth instants. This phenomenon 
is close to the Fermi-shuttle type ionization well know either in cosmology \cite{fermi} or in ion-atom collisions \cite{sulik}.

\section{Conclusions}

We presented both a full quantum mechanical and a semiclassical calculations of above threshold ionization 
of hydrogen atom at the tunneling regime by a few-cycle infrared laser pulse. 
We found that the 3 dimensional semiclassical method can describe reasonably well the momentum 
correlation pattern of the ATI peak. We also show that for multi-cycle pulse, good agreement between 
the results obtained by TDSE method and the semiclassical method can be reached. Furthermore, with the semiclassical approximation
we clearly identify and separate the regions in momentum distributions of the ejected electrons according to initial conditions. 
We illustrated the corresponding regions with typical electron trajectories.

\paragraph{Acknowledgments}
This work was supported by the Hungarian Scientific Research Fund OTKA Nos. NN 103279 and K103917, 
by the COST Actions CM1204 (XLIC) and CM1405 (MOLIM).


\begin{thebibliography}{00}


\bibitem{Ay}
\bibinfo{author}{Ayadi, V.}, \bibinfo{author}{Benedict M. G.},
 \bibinfo{author}{Dombi, P.} \& \bibinfo{author}{F\"oldi P.}
\newblock \url{http://arxiv.org/abs/1604.03437}.

\bibitem{Li} Li M., \emph{et~al.}, Phys Rev Lett. {\bf 112}, 113002 (2014)

\bibitem{Hu} Hu B., Liu, J., \& Chen, S. G., Phys. Lett. A {\bf 236}, 533 (1997) 

\bibitem{Fey} Feynman, R. P., Rev. Mod. Phys. {\bf 20}, 367 (1948)

\bibitem{BandraukPhysRevA04}
\bibinfo{author}{Chelkowski, S.}, \bibinfo{author}{Bandrauk, A.~D.} \&
  \bibinfo{author}{Apolonski, A.}
\newblock \emph{\bibinfo{journal}{Phys. Rev. A}} \textbf{\bibinfo{volume}{70}},
  \bibinfo{pages}{013815} (\bibinfo{year}{2004}).

\bibitem{Mourou}
\bibinfo{author}{Mourou, G. A.}, \bibinfo{author}{Tajima, T.} \& \bibinfo{author}{Bulanov, S. V.}
Rev. Mod. Phys. 78 309 (2006)

\bibitem{BandraukJPhysB13}
\bibinfo{author}{Bandrauk, A.~D.}, \bibinfo{author}{Fillion-Gourdeau, F.} \&
  \bibinfo{author}{Lorin, E.}
\newblock \emph{\bibinfo{journal}{J. Phys. B: At. Mol. Opt. Phys.}}
  \textbf{\bibinfo{volume}{46}}, \bibinfo{pages}{153001}
  (\bibinfo{year}{2013}).

\bibitem{ReissJPhysB14}
\bibinfo{author}{Reiss, H.~R.}
\newblock \emph{\bibinfo{journal}{J. Phys. B: At. Mol. Opt. Phys.}}
  \textbf{\bibinfo{volume}{47}}, \bibinfo{pages}{204006}
  (\bibinfo{year}{2014}).

\bibitem{Krause92}
\bibinfo{author}{Krause, J.~L.}, \bibinfo{author}{Schafer, K.~J.} \&
  \bibinfo{author}{Kulander, K.~C.}
\newblock \emph{\bibinfo{journal}{Phys. Rev. A}} \textbf{\bibinfo{volume}{45}},
  \bibinfo{pages}{4998--5010} (\bibinfo{year}{1992}).
  
\bibitem{MullerEfficient}
\bibinfo{author}{Muller, H.~G.}
\newblock \emph{\bibinfo{journal}{Laser Physics}} \textbf{\bibinfo{volume}{9}},
  \bibinfo{pages}{138 -- 148} (\bibinfo{year}{1999}).

\bibitem{PeacemanADI}
\bibinfo{author}{Peaceman, D.~W.} \& \bibinfo{author}{Rachford, H.~H.}
\newblock \emph{\bibinfo{journal}{Journal of the Society for Industrial and
  Applied Mathematics}} \textbf{\bibinfo{volume}{3}}, \bibinfo{pages}{28 -- 41}
  (\bibinfo{year}{1955}).

\bibitem{Bandrauk2005} Chelkowski, S. and Bandaruk, A. D.,  Phys. Rev. A. {\bf 71} 053815 (2005)

\bibitem{DombiOptExp05}
\bibinfo{author}{Dombi, P.} \emph{et~al.}
\newblock \emph{\bibinfo{journal}{Opt. Express}} \textbf{\bibinfo{volume}{13}},
  \bibinfo{pages}{10888--10894} (\bibinfo{year}{2005}).
\newblock \bibinfo{note}{And references therein}.    

\bibitem{Cavelieri}
\bibinfo{author}{Cavalieri, A.~L.} \emph{et~al.}
\newblock \emph{\bibinfo{journal}{New Journal of Physics}}
  \textbf{\bibinfo{volume}{9}}, \bibinfo{pages}{242} (\bibinfo{year}{2007}).

\bibitem{Stegun}
\bibinfo{author}{Abramowitz, M.} \& \bibinfo{author}{Stegun, I. A.}
\newblock \bibinfo{title}{Handbook of Mathematical Functions with Formulas, Graphs, and Mathematical Tables.}.

\bibitem{RKPD}
\bibinfo{author}{Dormand, J.R.} \& \bibinfo{author}{Prince, P.J.}
  (\bibinfo{year}{1980}).

\bibitem{Landau3} \bibinfo{author}{Landau, L. D.} \& \bibinfo{author}{Lifshitz, E. M.}
\newblock \bibinfo{title}{Quantum Mechanics: Non-Relativistic Theory. Vol. 3}.

\bibitem{Delone91}
\bibinfo{author}{Delone, N. B.} \& \bibinfo{author}{Krainov, V. P.}
\newblock \emph{\bibinfo{journal}{J. Opt. Soc. Am. B}}
  \textbf{\bibinfo{volume}{8}}, \bibinfo{pages}{1207} (\bibinfo{year}{1991}).

\bibitem{Delone98}
\bibinfo{author}{Delone, N. B.} \& \bibinfo{author}{Krainov, V. P.}
\newblock \emph{\bibinfo{journal}{Physics-Uspekhi}}
  \textbf{\bibinfo{volume}{41}}, \bibinfo{pages}{469} (\bibinfo{year}{1998}).

\bibitem{fermi} Fermi, E., Phys. Rev. {\bf 75}, 1169 (1949)

\bibitem{sulik} Sulik, B., Koncz, Cs., T\"ok\'esi, K., Orb\'an, A., Ber\'enyi, D., Physical Review Letters {\bf 88}, 073201 (2002)

  





\end{thebibliography}
\end{document}